\documentclass[fleqn,10pt]{article}
\usepackage[latin9]{inputenc}
\usepackage[a4paper]{geometry}
\usepackage{amsmath}
\usepackage{amssymb}
\usepackage{stackrel}
\usepackage{graphicx}
\usepackage{authblk}
\usepackage[unicode=true,
 bookmarks=false,
 breaklinks=false,pdfborder={0 0 1},backref=section,colorlinks=false]
 {hyperref}
\hypersetup{linkcolor=blue,urlcolor=blue,colorlinks,citecolor=blue}
\title{From Logistic Growth to Exponential Growth in a Population Dynamical
Model}
\begin{document}
\author[1]{Inbar Seroussi}
\author[2]{Nir Sochen}
\affil[1]{Department of Mathematics, Weizmann Institute of Science, POB 26, Rehovot 76100, Israel}
\affil[2]{Department of Applied Mathematics, School of Mathematical Sciences,
	Tel-Aviv University, Tel-Aviv, 69978, Israel}

\maketitle
\begin{abstract}
Dynamics among central sources (hubs) providing a resource and large
number of components enjoying and contributing to this resource describes
many real life situations. Modeling, controlling, and balancing this
dynamics is a general problem that arises in many scientific disciplines.
We analyze a stochastic dynamical system exhibiting this dynamics with
a multiplicative noise. We show that this model can be solved exactly
by passing to variables that describe the mass ratio between the components
and the hub. We derive a deterministic equation for the average
mass ratio. This equation describes logistic growth. We derive the
full phase diagram of the model and identify three regimes by calculating
the sample and moment Lyapunov exponent of the system. The first regime
describes full balance between the non-hub components and the hub, in
the second regime the entire resource is concentrated mainly in the hub,
and in the third regime the resource is localized on a few non-hub components and the hub. Surprisingly, in the limit of large number of components
the transition values do not depend on the amount of resource given
by the hub. This model has interesting application in the context
of analysis of porous media using Magnetic Resonance (MR) techniques. 
\end{abstract}

\section{Introduction}

Population dynamics on large scale networks has attracted a lot of
attention due to its wide occurrence in many disciplines, such as
social sciences \cite{Bouchaud2000536,castellano2009statistical},
physics \cite{kardar1986dynamic} and biology, communication and control
theory \cite{allegra2016phase}. This dynamics is mainly affected
by the topology of the network as well as some internal stochastic noise.
In many applications there are only a few nodes playing a major role
in the dynamical process, distributing and carrying most of the resources \cite{barabasi2000scale,song2005self}.
For example, this can be the case in models describing population dynamics, economic growth \cite{bouchaud2015growth},
and distributed control systems \cite{allegra2016phase}. An additional
application of this problem is in the context of diffusion measurements
of porous systems, such as brain tissue, using Magnetic Resonance
(MR) techniques. In this last case, the sensitivity of the MR signal to
self-diffusion of water molecules can be utilized to extract information
about the network of cells (neurons) in the brain. The concept of
self-diffusion of molecules in a network of pores was already introduced
in Refs. \cite{seroussi2018spectral,callaghan1992diffusion,callaghan2011translational}.
The main challenge is how to determine the topology of the network based
on the MR measurements \cite{seroussi2018spectral}.

Our model consists of a system of interacting sites on a graph $\mathcal{G}$
with $N$ vertices and $E$ edges between them. We are interested
in the stochastic dynamics of some characteristic property $\{m_{i}(t)\}_{i\in\mathcal{G},t\text{\ensuremath{\ge}}0}$.
The property $m_{i}(t)$ is linked to a physical measurable quantity
in the real world and the graph is the underlying geometry/topology
in which the property lives. The topology is a complex network of
sites. The model is described by the following family of stochastic differential
equations in the Stratonovich form on the graph $\mathcal{G}$:

\begin{equation}
\frac{dm_{i}(t)}{dt}=\underset{j}{\sum}J_{ij}m_{j}(t)-\underset{j}{\sum}J_{ij}m_{i}(t)+g_{i}(t)m_{i}(t),\label{eq:General model}
\end{equation}
with the initial conditions $m_{i}(0)=m_{0}$. The term $g_{i}(t)$
is a multiplicative white noise, such that $\langle g_{i}(t)\rangle=f_{i}$,
and $\langle g_{i}(t)g_{j}(t')\rangle=\sigma_{i}^{2}\delta_{ij}\delta(t-t')$.
We choose the Stratonovich form, since its solution is a limiting
case of a physical system involving white noise with short memory \cite{van1981ito}.
The topology of the network is encoded in the adjacency matrix $J$
of the graph. The model consists of two parts: an interacting part,
where the interaction strength depends on the
location on the graph, and a non-interacting part, where each component
follows a stochastic noise with different variance $\sigma_{i}^{2}$.
The first part causes spreading, while the second pushes towards concentration
(a.k.a localization or condensation). The model was already analyzed
in the mean field topology, i.e., when all the nodes are connected and interact
at the same rate. In this case, the equilibrium distribution is a
Pareto power-law \cite{Bouchaud2000536}. It was also analyzed on
trees \cite{derrida1988polymers,gueudre2014explore}, and random graphs
assuming separable probability distribution on the nodes \cite{ichinomiya2012bouchaud}.
The model on the lattice is known in the mathematical literature as
the time-dependent Parabolic Anderson Model (PAM) \cite{carmona1994parabolic,molchanov1991ideas}.
The phase diagram of the model in this case depends on the dimension
of the lattice. On a general network, phase transitions depend
on the spectral dimension of the network \cite{seroussi2018spectral}.

Here, we present and analyze a specific topology in which the model
is shown to be solvable. Namely, we consider a directed graph with $N+1$ nodes, one hub
node interacting with $N$ independent nodes. In the context of MR
measurements of diffusion in a porous structure, the MR signal measured
is assumed to be composed of two contributions: one coming from hindered
diffusion in the extracellular space and the other from restricted
diffusion in the intracellular space \cite{karger1985nmr,moutal2018karger,niendorf1996biexponential,mulkern1999multi}.
The hub node represents the magnetization in the extracellular space (e.g.,
water), $h_{0}(t)$, and the non-hub nodes represents $N$ independent
intracellular pores with magnetization, $m_{i}(t)$. The motion of molecules between these regions changes the value of the magnetization
as a function of time and is represented by the interaction term between
nodes, $J_{ij}$. The effect of the magnetic field gradient can be
incorporated in the stochastic noise, for example, in $f_{i},$ and/or
its variance $\sigma_{i}^{2}$. In the economic context, the system
describes the dynamics of the money hold by the hub, which represents by
the state/bank, and the money of each agent $m_{i}(t)$. In this case,
the agents deposit money in the bank and the bank pays interest on it. The stochastic noise represents the bank/state and the agent's investments in the stock market and housing \cite{Bouchaud2000536,bouchaud2015growth}. Analysis of the dynamics of the sums of the money held by
the agents and the bank/state was curried out in Ref. \cite{bouchaud2015growth}.

Here, we analyze the dynamics of the mass ratio between each agent
and the bank/state (hub). We derive the equilibrium distribution in this
case, and show that when the number of nodes growth at least exponentially with time there exist a localization phase. 
We also generalize our analysis to multiple
number of hubs.

Our main result is a full phase-diagram of the model. We show that
this model can be described by a stochastic equation for the mass ratio
between each of the non-hub nodes and the hub, and a deterministic non-linear equation for the average relative mass of all the non-hub nodes
with respect to the hub. 
To identify the phases of the system, we calculate the sample and
moment Lyapunov exponents and identify a gap between them. The phase
transitions are characterized by one parameter. This parameter takes
into account the exchange rate between the non-hub nodes and the hub
and the variances of the multiplicative noises. 

\section{Hub Topology\label{sec:Hub-Topology}}

\begin{figure}
\centering{}\center{\includegraphics[trim={3cm 2cm 4cm 2cm},clip,scale=0.4]{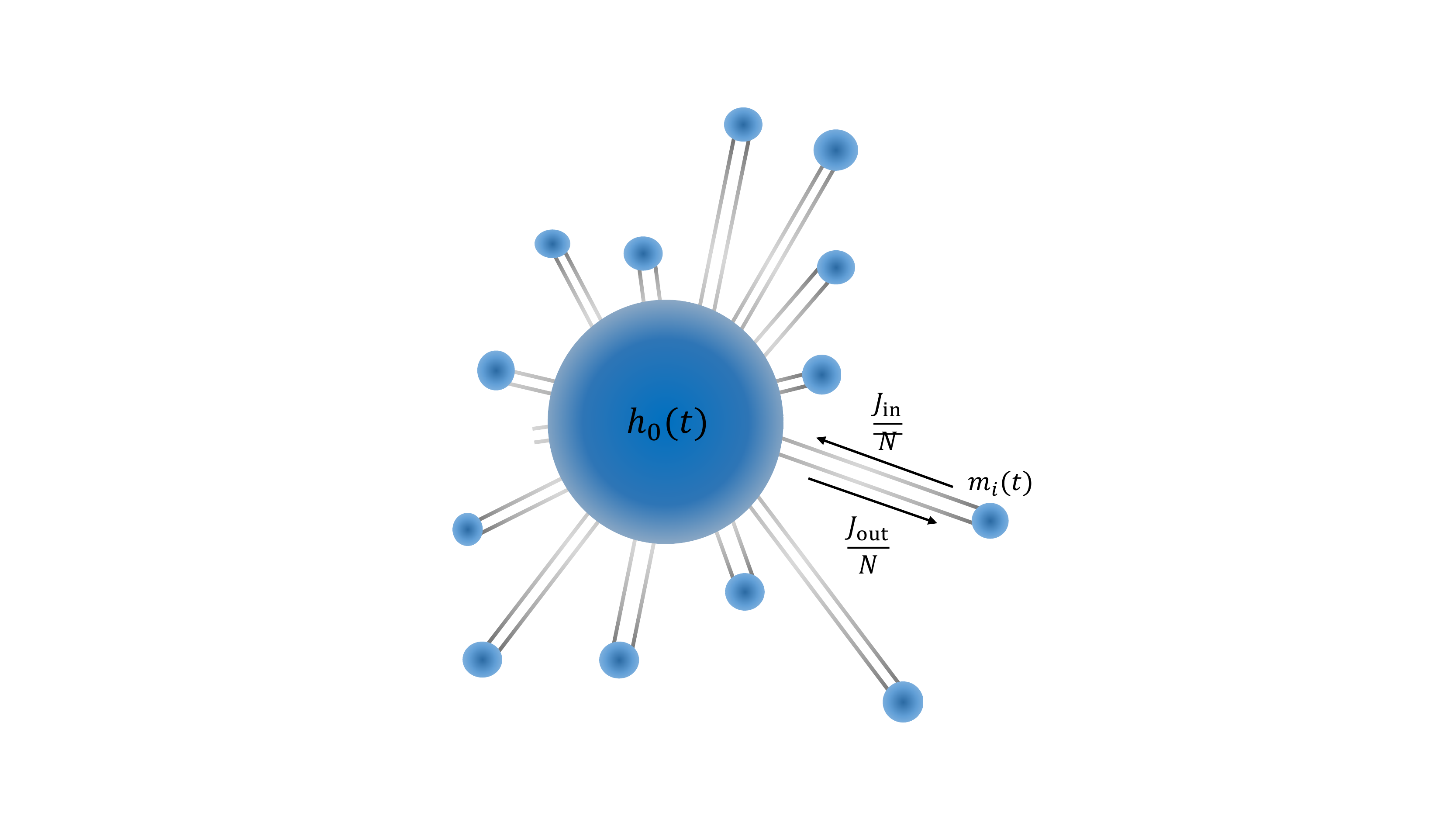}\caption{An illustration of the system in the hub topology.\label{fig:Hub-Topology}}
} 
\end{figure}

The basic hub topology is composed of an infinite number of nodes, $\{m_{i}\}$,
interacting at constant rate with a hub node, $h_{0}$, such that,
$J_{i0}=\frac{J_{\mathrm{out}}}{N}$, and $J_{0j}=\frac{J_{0}}{N}$,
respectively. Our normalization is such that, the overall interaction
between the nodes and the hub is finite in the limit of infinite
number of nodes. The interaction among the non-hub nodes is characterized by
the parameter $\delta$; when $\delta=0$, any interaction (transfer
of mass) between the non-hub nodes is done only through the hub. The
topology of the interaction between the non-hub nodes is defined by
a Laplacian matrix, $L$, satisfying $\sum_{i}L_{ij}=0$. Figure \ref{fig:Hub-Topology}
presents an illustration of such a system for $\delta=0$. 

We assume that the stochastic noise acting on the non-hub nodes has the
same variance and average for all nodes, $\sigma_{i}=\sigma_{\mathrm{out}}$, $f_{i}=f$.
Eq. (\ref{eq:General model}) in the Itô form  reduces to the following
system of stochastic equations:

\begin{equation}
\frac{dh_{0}}{dt}=\frac{J_{\mathrm{in}}}{N}\underset{j}{\sum}m_{j}-J_{\mathrm{out}}h_{0}+f_{0}h_{0}+\frac{\sigma_{0}^{2}}{2}h_{0}+\sigma_{0}g_{0}h_{0},\label{eq:m0 Ito}
\end{equation}

\begin{equation}
\frac{dm_{i}}{dt}=\frac{J_{\mathrm{out}}}{N}h_{0}-\frac{J_{\mathrm{in}}}{N}m_{i}+fm_{i}+\frac{\sigma_{\mathrm{out}}^{2}}{2}m_{i}-\delta\underset{j}{\sum}L_{ij}m_{j}+\sigma_{\mathrm{out}}g_{i}m_{i}.\label{eq:mi Ito}
\end{equation}
The Itô form will be useful latter on when one takes the limit $N\rightarrow\infty$.
It is instructive to pass to the following normalized variables by
introducing the mass ratio between the non-hub nodes and the hub node,
$M_{i}=\frac{m_{i}}{h_{0}}$. This leads to the following system of
$N$ equations in the Itô form (See \ref{sec:Variable-transformations}
for more details): 
\begin{equation}
\frac{dM_{i}}{dt}=\frac{J_{\mathrm{out}}}{N}-\left(\frac{J_{\mathrm{in}}}{N}+\Delta f-J_{\mathrm{out}}-\frac{\sigma^{2}}{2}+J_{\mathrm{in}}\overline{M}(t)\right)M_{i}-\delta\underset{j}{\sum}L_{ij}M_{j}+\sigma\xi_{i}M_{i}.\label{eq:stochastic Mi}
\end{equation}
We introduce the average mass ratio $\overline{M}=\frac{1}{N}\sum_{i}M_{i}$,
the effective variance $\frac{\sigma^{2}}{2}=\frac{\sigma_{0}^{2}+\sigma_{\mathrm{out}}^{2}}{2}$,
and the average difference $\Delta f=f_{0}-f$. Here, $\xi_{i}(t)$
is a Gaussian process with mean zero and variance one. The 
transformation of variables introduces a non-linear term, which accounts for the interaction
between any two nodes through the hub. In the limit  $N\rightarrow\infty$,
averaging over all the nodes in Eq. (\ref{eq:stochastic Mi}) yields
the following deterministic law for the average mass ratio: 
\begin{equation}
\frac{d\overline{M}}{dt}=\frac{\sigma^{2}\beta}{2}\overline{M}\left(\frac{\alpha+1}{\beta}-\overline{M}\right),\label{eq:Mbar deterministic}
\end{equation}
where we use the dimensionless parameters $\alpha=2\frac{J_{\mathrm{out}}-\Delta f}{\sigma^{2}}$
and $\beta=\frac{2J_{\mathrm{in}}}{\sigma^{2}}$. Since, Eq. (\ref{eq:Mbar deterministic}) it is a non-linear equation, there are two steady-state solutions: $\overline{M}_{1\mathrm{eq}}{\rightarrow}\frac{J_{\mathrm{out}}-\Delta f+\frac{\sigma^{2}}{2}}{J_{\mathrm{in}}}=\frac{\alpha+1}{\beta}$ as $N\rightarrow\infty$,
and $\overline{M}_{2\mathrm{eq}}{\rightarrow}0$ as $N\rightarrow\infty$.
Note that, in the case of $\delta=0$ the system geometry can be viewed
as a directed tree structure with one level and infinitely many leaves. Therefore, the presence of a the non-linear term is caused by
to the indirect interaction between the non-hub nodes following the tree
topology of the system \cite{derrida1988polymers}. Convergence to
each one of these fixed points depend on the initial condition of the system
and on the stability of these points. Stability analysis of these
two points shows that the point $\overline{M}_{1\mathrm{eq}}$ is
stable when $\alpha+1>0$, while the point $\overline{M}_{2\mathrm{eq}}=0$,
is stable when $\alpha+1\leq0$. For example, in the context of MR
measurements in porous media, this system can model a complex structure
measured from a single voxel in the MR image. The value $\overline{M}_{\mathrm{eq}}$
describes the steady-state average magnetization ratio, between the
extracellular space and the intracellular space. The first fixed point
$\overline{M}_{1\mathrm{eq}}$ is reached when the steady-state
magnetization ratio is equal to the amount of molecules leaving the
non-hub pores reduced by the magnetic field effects divided by the amount
of molecules leaving the extracellular space. The second fixed point
$\overline{M}_{2\mathrm{eq}}$ represents the case in which on average most
of the contribution to the magnetization in a single voxel comes mainly
from the extracellular space. Eq. (\ref{eq:Mbar deterministic}) shows
logistic growth and is a version of Lotka-Volterra equation, which describes
many social phenomena in nature \cite{lotka1925elements,volterra1926fluctuations}.
It can be solved exactly: for an initial condition $\overline{M}(0)=\overline{M}_{0}$, the solution is
\begin{equation}
\overline{M}(t)=\frac{\overline{M}_{0}(\alpha+1)}{\beta\left(\overline{M}_{0}+\left(\frac{\alpha+1}{\beta}-\overline{M}_{0}\right)e^{-\frac{\sigma^{2}\left(\alpha+1\right)t}{2}}\right)}.\label{eq:Mbarsol}
\end{equation}

\subsection{Equilibrium Distribution\label{subsec:Equilibrium-Distribution}}

Given Eqs. (\ref{eq:stochastic Mi}) and  (\ref{eq:Mbar deterministic})
for the relative mass between the non-hub nodes and the hub, one can
derive an equivalent form describing the dynamics of the probability
distributions of $M_{i}$. Since the dynamics of the average mass
ratio between the non-hub nodes and the hub node is governed by a
deterministic non-linear equation, in the limit
$N\rightarrow\infty$ the system reduces to a set of stochastic independent equations
for the relative mass in each node, $M_{i}$. Therefore, we can omit
the index $i$, and look at the dynamics of the probability distribution
of a typical node $P(M,t)$. This dynamics of the probability distribution
is described by the following Fokker-Plank equation:

\begin{equation}
\frac{\partial P}{\partial t}=-\frac{\sigma^{2}}{2}\frac{\partial\left(\left(\left(\alpha+1-\beta\,\overline{M}-\tilde{\delta}\right)M+\tilde{\delta}\,\overline{M}(t)-\tilde{\delta}\right)P\right)}{\partial M}+\frac{\sigma^{2}}{2}\frac{\partial^{2}\left(M^{2}P\right)}{\partial^{2}M}.\label{eq:FP M0}
\end{equation}
To study the effect of interaction between nodes, we introduce a small mean field interaction between
the non-hub nodes, $\delta$, and define the dimensionless interaction rate,
$\tilde{\delta}=\frac{2\delta}{\sigma^{2}}$. 
By equating the left hand side to of Eq. (\ref{eq:FP M0}) zero one can find the steady-state distribution.
The solution shows a Pareto power-law behavior: 
\begin{equation}
P_{\mathrm{eq}}(M)=A\,\mathrm{exp}\left(-\frac{\tilde{\delta}\,\overline{M}_{\mathrm{eq}}}{M}\right)M^{-\mu(\overline{M}_{\mathrm{eq}})}.
\end{equation}
The power is a function of the steady-state average relative mass
$\overline{M}_{\mathrm{eq}}$, i.e., steady-state solution of Eq.
(\ref{eq:Mbar deterministic}): $\mu(\overline{M}_{\mathrm{eq}})=\beta\overline{M}_{\mathrm{eq}}+1-\alpha+2\tilde{\delta}$.
Substituting the value of the average mass ratio, we find that \textbf{$\mu(\frac{\alpha+1}{\beta})|_{\alpha+1\geq0}=2+2\tilde{\delta}$},
and $\mu(0)|_{\alpha+1<0}\rightarrow1-\alpha+2\tilde{\delta}$. This
shows that the system has two steady-states. The system collapses
to one of them depending on the initial condition, i.e., the initial
mass ratio between the hub and the non-hub nodes. Note that the value
of $\mu$ is greater than $2$ when the system collapses to the state $\overline{M}_{\mathrm{eq}}=\frac{\alpha+1}{\beta}$,
showing equality among the non-hub nodes and the hub. For, $\overline{M}_{\mathrm{eq}}=0$,
the mass is localized on a few nodes within the set of non-hub nodes. The power is, $\mu<2$, as long
as $2\tilde{\delta}<1+\alpha$. Therefore, adding interaction between non-hub nodes reduces localization, as expected.

\subsection{Balance and Localization\label{sec:Balance-and-Localization}}

The analysis above reveals the localization regime within the non-hub nodes when the influence of the hub is renormalized. In this section, we analyze the regime at  which there is localization on the hub. For this purpose, we study the asymptotic
properties of the total mass, $E(t)=h_{0}(t)+\sum_{i}m_{i}(t)$. We calculate the Lyapunov exponents of the solution \cite{carmona1994parabolic,molchanov1991ideas}.
Here, we perform the analysis for the case of $\delta=0.$ The Lyapunov
exponents describe the growth rates of the solution and its moments. They indicate the level of complexity in the solution's landscape.
The first moment Lyapunov exponent of the solution is as follows:
\begin{equation}
\gamma_{1}=\underset{t,N\rightarrow\infty}{\mathrm{lim}}\,\frac{1}{t}\mathrm{ln}\left(\langle E(t)\rangle\right)=\begin{cases}
\begin{array}{c}
f+\frac{\sigma_{\mathrm{out}}^{2}}{2}\\
f_{0}-J_{\mathrm{out}}+\frac{\sigma_{0}^{2}}{2}
\end{array} & \begin{array}{c}
\frac{\Delta\sigma^{2}}{\sigma^{2}}<\alpha\\
\frac{\Delta\sigma^{2}}{\sigma^{2}}\geq\alpha
\end{array}\end{cases},\label{eq:gamma1}
\end{equation}
where $\Delta\sigma^{2}=\sigma_{0}^{2}-\sigma_{\mathrm{out}}^{2}$ is the variance difference,
see \ref{sec:Moments-equations} for details of the proof together
with corrections for finite network size. Interestingly, what determines
the growth on average is the difference between the variance, $\Delta\sigma^{2}$
of the stochastic noises of the hub and that of the non-hub nodes. To understand Eq. (\ref{eq:gamma1}), let us look at the limit where
the stochastic noise in the non-hub nodes has significantly higher variance
compared to the hub. This is equivalent to the presence of large fluctuations in
the non-hub pores with respect to the extracellular space, i.e., $\Delta\sigma^{2}\approx-\sigma_{\mathrm{out}}^{2}$.
In this case, comparing to the stability points in Eq. ($\ref{eq:Mbar deterministic}$)
in the regime $\alpha\leq-1$, there is a high concentration of magnetization on the hub and a few non-hub nodes which contribute most to the total growth rate.
For $\alpha>-1$, in the detailed balance limit the non-hub nodes and the hub contribute equally to the total growth.
On the other hand, in the limit of $\Delta\sigma^{2}\approx\sigma_{0}^{2}$,
the system exhibits three regimes: for $\alpha>1$ the magnetization spreads equally among the nodes, for $-1<\alpha\leq1$ the magnetization
is mainly concentrated on the hub, and for $\alpha<-1$,
there is concentration of the magnetization on the hub and/or several non-hub nodes. This analysis is verified by calculating the value of the sample Lyapunov exponent, which provide the sample
growth rate. 
 
We define the
sample Lyapunov exponent, $\widetilde{\gamma}$, as the limit of the
logarithm of the total mass of the solution divided by time as $N,t\rightarrow\infty$. Knowing the dynamics of the average mass ratio, see Eq.
(\ref{eq:Mbar deterministic}), we can calculate the sample growth
rate of the mass ratio exactly. The resulting sample/quenched Lyapunov
exponent is 
\begin{equation}
\widetilde{\gamma}=\underset{t,N\rightarrow\infty}{\mathrm{lim}}\,\frac{1}{t}\mathrm{ln}\left(h_{0}+\overline{m}\right)=f+\frac{\sigma_{\mathrm{out}}^{2}}{2},\label{eq:sample Lyapunov}
\end{equation}
where we denote $\overline{m}=\sum_{i}m_{i}$; see \ref{sec:Sample-Lyapunov-Exponent}
for details of the proof. Note that, in order to take the limit one needs to specify at which rate the number of nodes growth with time. We show that when the number of nodes growth at least exponentially with time, then the sample Lypunov is as in Eq. (\ref{eq:sample Lyapunov}). 
This value is independent of the initial conditions and is bounded
from above by the first moment Lyapunov exponent, $\gamma_{1}$. Localization
of the solution is defined as the regime at which strict
inequality hold, $\widetilde{\gamma}<\gamma_{1}$ \cite{carmona1994parabolic}.
The gap $\Delta\gamma=\gamma_{1}-\widetilde{\gamma}$ between these two exponents
i.e., the difference between the expression in Eq. (\ref{eq:gamma1}) and Eq. (\ref{eq:sample Lyapunov}),
characterizes the localization regime
in the system. Combining the transition in the values of the exponent $\widetilde{\gamma}$
with the stability analysis of the steady-state solutions
of Eq. (\ref{eq:Mbar deterministic}),
we identify three regimes: a regime of full equality,
for $\alpha>\frac{\Delta\sigma^{2}}{\sigma^{2}}$, in which the mass is spread
equally between all the non-hub nodes and the hub, a second regime
for $-1<\alpha\leq\frac{\Delta\sigma^{2}}{\sigma^{2}},$ in which the
mass is localized mainly on the hub, and a third regime for
$\alpha\leq-1$, in which 
the mass is localized on the hub and a few non-hub nodes. 

\section{Multiple Hubs Topology \label{Sec: multipileHaubs}}

\begin{figure}
\begin{centering}
\includegraphics[trim={3cm 2cm 3cm 3cm},clip,scale=0.3]{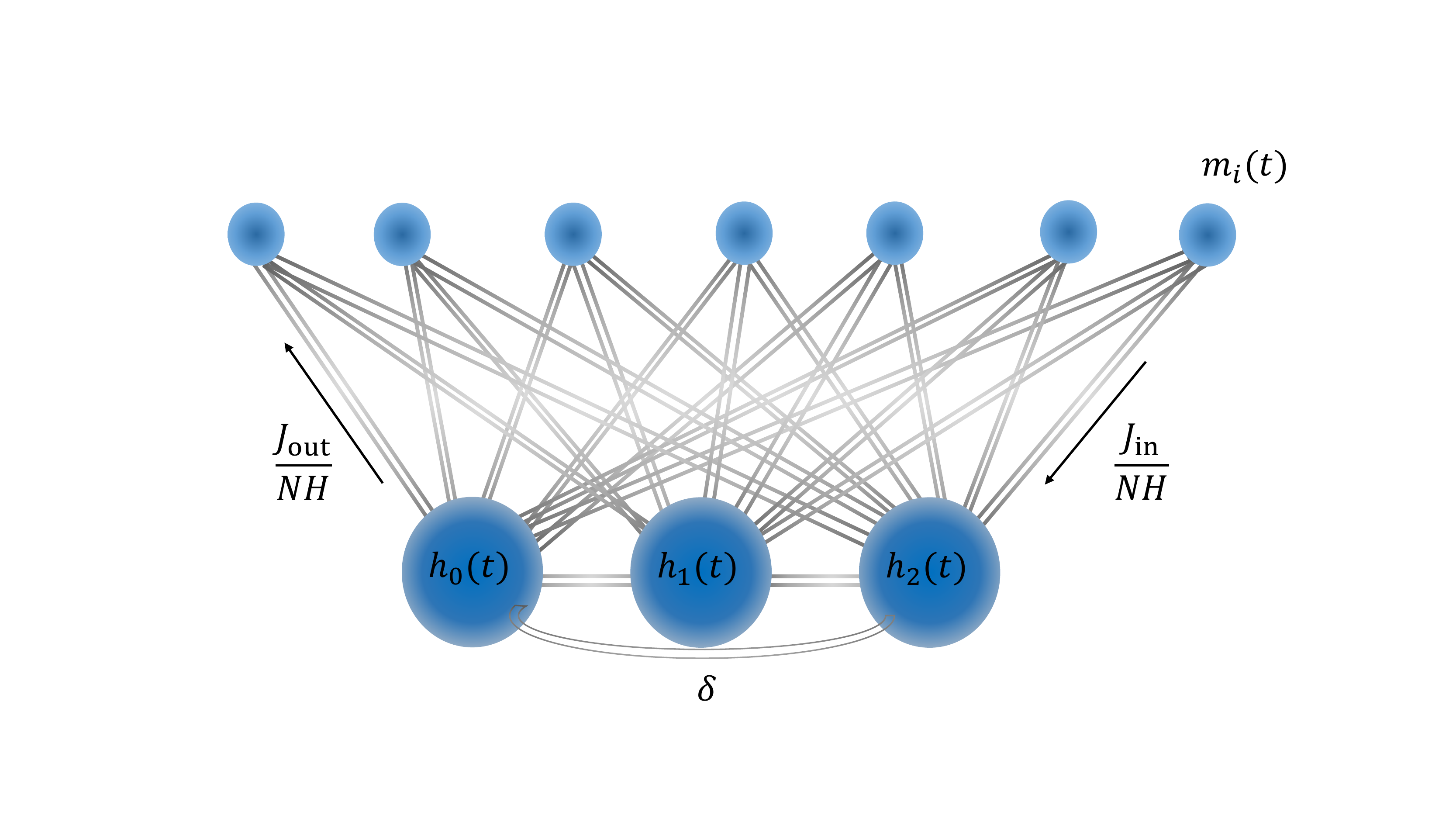} 
\par\end{centering}
\caption{An illustration of the multi-hub topology for $M=3$, and, $N=7$.\label{fig:Multipile hubs} }
\end{figure}

In this section, we consider the effect of $H$ hub nodes, $h_{i},$
connected to all the nodes in the system, and a set of $N$ non-hub
independent nodes, $m_{i}$, connected only to the hubs nodes. Figure \ref{fig:Multipile hubs}
illustrate of this topology. This kind of topology appears in
many applications, for instance, in the economic setting in the presence of more
than one central bank/company. In a porous structure, it can describe different
extra-cellular regions interacting with intracellular pores. It is
also applied in analyzing the dynamics of control systems \cite{allegra2016phase},
and in machine learning algorithms. The equations of the system now read
as follows: 
\begin{equation}
\frac{dh_{i}}{dt}=\frac{J_{\mathrm{in}}}{HN}\sum_{i}m_{i}-\frac{J_{\mathrm{out}}}{H}h_{i}+f_{i}h_{i}+\frac{\delta}{H-1}\sum_{j\neq i}^{H}h_{j}-\delta h_{i}+\sigma_{i}g^h_{i}h_{i},\label{eq:hi-multi-hub}
\end{equation}

\begin{equation}
\frac{dm_{i}}{dt}=\frac{J_{\mathrm{out}}}{NH}\sum_{i}h_{i}-\frac{J_{\mathrm{in}}}{N}m_{i}+q_{i}m_{i}+\nu_{i}g^m_{i}m_{i}.\label{eq:mi-multi-hub}
\end{equation}
Note that, as before, the total interaction rates between the hubs
and the non-hub nodes, $J_{\mathrm{in}},$ and $J_{\mathrm{out}}$,
are defined to be finite in the limit of infinite $N$ and $H$.
The processes $g^h_{i}(t)$ and $g^m_{i}(t)$, are Gaussian processes
with mean zero and variance one. Similar to the one-hub topology,
we can now pass to the relative mass parameters by dividing by
the average hubs mass, see \ref{sec:Multiple-Hubs-Derivation} for
more details. The results of the previous sections are recovered when
$H=1$. Note that the equations for the relative mass are decoupled
in the case $H=1$ and also in the limit of $H$ very large. In
the presence of a finite small number of hub nodes, one can show that
the total variance depends on the hubs value. Therefore, having finite
number of hubs decreases the value of the parameter $\alpha$, and
causes more equality in the system and reduces the localization. Moreover,
in the simple case where all the hubs have the same statistics, such
that the stochastic noise, $g^h_{i}(t)=g(t)$, does not depend on the
hub location, $H$ hubs are equivalent to one hub with the total effective
net flux $J_{\mathrm{out}}$. The limit of one non-hub node
with $J_{\mathrm{out}}=J_{\mathrm{in}}=\delta$, $N=1$, and an infinite
number of hubs, $H\rightarrow\infty$, is the mean field model with
exponential growth of the total mass \cite{Bouchaud2000536}. Note
that when both the number of non-hub nodes and the number of hubs are
very large, i.e., $N\rightarrow\infty$ and $H\rightarrow\infty$, the
average mass ratio grows exponentially. The exponent depends on the
average and variance difference of the stochastic multiplicative noises.
The phase transition in this case is equivalent to the results in Sec. \ref{sec:Hub-Topology},
i.e., there are three phases: localization on the non-hub nodes, localization on the hubs and equal spreading over all the nodes. 
Note that in this limit, the non-hub nodes play the same role as the
hubs, since they are connected to infinitely many  hub nodes. The interaction among the hubs, $\delta$ reduces localization, but
doesn't effect the growth rate. The analysis above affects mainly 
the sample Lyapunov exponent. The first moment Lyapunov exponent remains
the same for any $H$ and $N$, since the average equations do not
change. This shows that the phase transitions predicted in Sec. \ref{sec:Balance-and-Localization}
are general and can be observed with small modifications to a system
of multiple hubs. In addition, the transition between a logistic growth
in the relative mass to an exponential growth is a function of the ratio between the $N$ and $H$. 

\section{A Note on MRI}

\begin{figure}
\begin{centering}
(a)\includegraphics[trim={6cm 10cm 6cm 11cm},clip,scale=0.68]{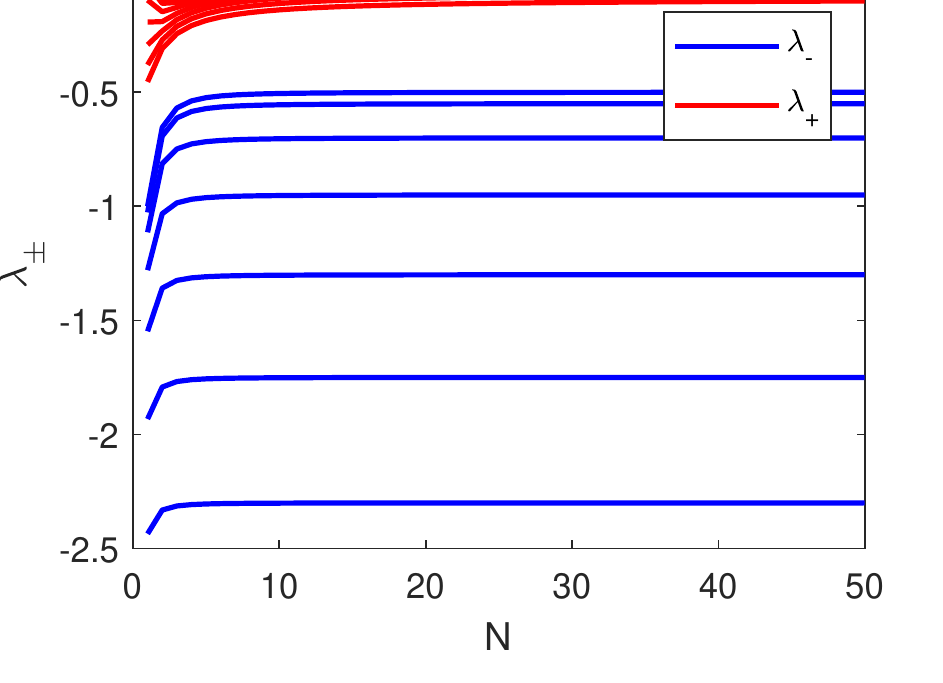}(b)\includegraphics[trim={6cm 10cm 6cm 11cm},clip,scale=0.68]{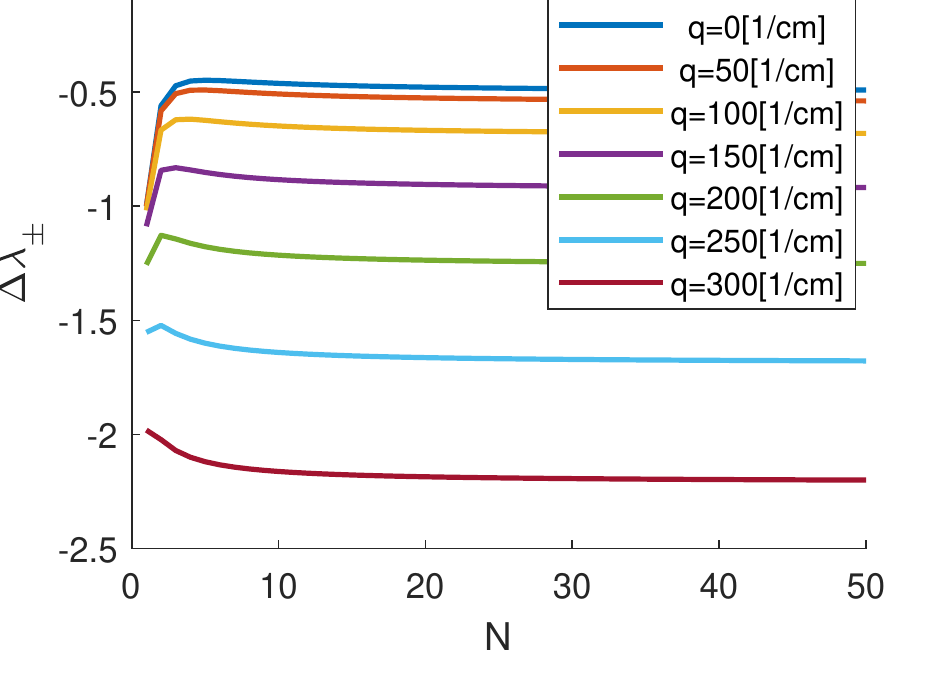} 
\par\end{centering}
\caption{(a) The eigenvalue as a function of the number of independent pores
for different values of the wave vector $q$ (b) The eigenvalue difference as a function of the number of independent pores for different
values of the wave vector $q$.\label{fig:The-eigenvalue-vsN}}
\end{figure}

In the context of MR measurements the model in Eq. (\ref{eq:General model})
is a generalization of the Kärger model \cite{karger1985nmr}, which
accounts for random changes in the diffusivity due to restricted diffusion
or a non-homogeneous magnetic field. This model was already analyzed
on a general network, where the importance of the spectral dimension
as a measurable parameter is stated \cite{seroussi2018spectral}.
The hub topology (see Figure \ref{fig:Hub-Topology}) is a simplified
version of this model. We show that in this topology under the assumption
that all the non-hub pores have similar properties, the average equations
of the model are those for the Kärger model for two compartments \cite{karger1985nmr},
see \ref{sec:Moments-equations}. The parameters are then as follows:
$f_{0}=-q^{2}D_{\mathrm{ex}}$ and $f=-q^{2}D_{\mathrm{in}}$, $\sigma_{0}^{2}=-q^{2}\sigma_{\mathrm{ex}}^{2}$,
$\sigma_{\mathrm{out}}^{2}=-q^{2}\sigma_{\mathrm{in}}^{2}$, such
that the parameter $\alpha(q^{2})=2\frac{J_{\mathrm{out}}-\Delta f}{\sigma^{2}}=-2\frac{J_{\mathrm{out}}+q^{2}\left(D_{\mathrm{in}}-D_{\mathrm{ex}}\right)}{q^{2}\left(\sigma_{\mathrm{ex}}^{2}+\sigma_{\mathrm{in}}^{2}\right)}$,
is controlled by the gradient of the applied magnetic field, incorporated
in the value of $q$. For example consider the basic Stejskal-Tanner sequence \cite{stejskal1965spin},
which is composed of two gradients pulses of the magnetic field with
magnitude $G$ in opposite direction and with duration $\delta$.
The pulses are separated by a diffusion time $\Delta$. For
$q$ one takes the wave vector, $q=\frac{\delta\gamma G}{2\pi}$, where the parameter, $\gamma$, is the gyro-magnetic ratio. The average
equations for the magnetization in the $(x,y)-$plane, and with a magnetic
field gradient in the $\hat{z}$ direction, read

\begin{equation}
\frac{d\langle h_{0}\rangle}{dt}=\frac{J_{\mathrm{in}}}{N}\langle\overline{m}\rangle-J_{\mathrm{out}}\langle h_{0}\rangle-q^{2}D_{\mathrm{ex}}\langle h_{0}\rangle,\label{eq:m0 Ito-MRI params}
\end{equation}

\begin{equation}
\frac{d\langle\overline{m}\rangle}{dt}=\frac{J_{\mathrm{out}}}{N}\langle h_{0}\rangle-\left(\frac{J_{\mathrm{in}}}{N}+q^{2}D_{\mathrm{in}}+\frac{q^{2}\sigma_{\mathrm{in}}^{2}}{2}\right)\langle\overline{m}\rangle.\label{eq:mi Ito-MRI params}
\end{equation}
Note that, the wave vector $q$ turns on the stochastic dynamics.
We denote $\delta D=D_{\mathrm{ex}}-D_{\mathrm{in}}$, and $\delta\sigma^{2}=\sigma_{\mathrm{ex}}^{2}-\sigma_{\mathrm{in}}^{2}$.
The multiplicative white noise accounts for the random diffusivity
changes of the medium due to restricted diffusion. Based on the analysis
in Sec. \ref{sec:Balance-and-Localization}, one can find the transition
point in terms of the wave vector $q$. The transition point is at
$q_{\mathrm{c}}=\sqrt{\frac{J_{0}}{\delta D-\frac{\delta\sigma^{2}}{2}}}$.
The average signal reveals only the transition at $q_{\mathrm{c}}$. Note that,
the signal decay is also affected by the noise variance of the extracellular
space. Taking typical values such as, $J_{\mathrm{out}}=\frac{1}{1800\mathrm{msec}}=0.5556[\frac{1}{\mathrm{sec}}]$,
$D_{\mathrm{ex}}=2e-5[\frac{\mathrm{cm}^{2}}{\mathrm{sec}}]$, $D_{\mathrm{in}}=0.1e-5[\frac{cm^{2}}{\mathrm{sec}}]$.
Then the critical value is $q_{\mathrm{c}}\approx\sqrt{\frac{J_{\mathrm{out}}}{D_{\mathrm{ex}}-D_{\mathrm{in}}+\frac{\sigma_{\mathrm{in}}^{2}}{2}}}=\sqrt{\frac{0.5556}{1.9e-5}}\approx171[\frac{1}{\mathrm{cm}}]=0.1[\frac{1}{\mathrm{\mu m}}]$.
A larger variance in the non-hub pores will set the transition for a lower value of
$q$, value. Whereas larger variance in the diffusion in the extracellular
space will set the transition to a higher value of $q$ value. The decay of the
signal has a bi-exponential form as predicted by the Kärger model:

\begin{equation}
\gamma_{1}=\underset{t,N\rightarrow\infty}{\mathrm{lim}}\,\frac{1}{t}\mathrm{ln}\left(\langle E(t)\rangle\right)=\begin{cases}
\begin{array}{c}
-q^{2}\left(D_{\mathrm{in}}+\frac{\sigma_{\mathrm{in}}^{2}}{2}\right)\\
-q^{2}D_{\mathrm{ex}}-J_{\mathrm{out}}
\end{array} & \begin{array}{c}
q>q_{\mathrm{c}}\\
q\leq q_{\mathrm{c}}
\end{array}\end{cases}.
\end{equation}
The stochastic model is a natural generalization of the Kärger model.
Using this generalization, we are able to explore and analyze the
behavior of the model in the presence of complex topological structures, as well as the effect of changes in the apparent diffusivity as a
result of stochastic noise. Note that this transition appears also in the presence of any interaction $\delta$ among the non-hub nodes.
Figure \ref{fig:The-eigenvalue-vsN} presents the behavior of the eigenvalues
as a function of $N$. 

\section{Discussion and Conclusion }

We have presented a stochastic model that describes  diffusion on a graph
with an additional multiplicative stochastic noise. We analyze this
model on a directed graph with one hub node that is connected to a large
number of non-hub nodes. We derive a non-linear equation for the average
mass ratio between the non-hub nodes and the hub. This equation describes
logistic growth. It has two phases one in which the overall mass
is mainly concentrated on the hub, and the other in which there is a ``detailed balance'' such that the steady-state depends on the ratio
between the exchange rate between the hub and the non-hub nodes. We
show that this model is completely solvable in the large $N$ limit.
In addition, we identify the phase transitions of the model in terms of
the two parameters $\alpha$ and $\frac{\Delta\sigma^{2}}{\sigma^{2}}$. We show that in order for localization phase occur the number of non-hub nodes needs to grow at least exponentially with time.  
Surprisingly, in the limit of large number of independent
nodes the transition points do not depend on the amount of resource given
by the hub, provided that it is finite and non-zero. We generalize
this analysis to a system of multiple hubs. We show that in the limit
of infinitely many hubs the growth of the system becomes exponential.

The model has numerous applications. We introduce an application of
this model in the context of MR measurements of complex structures.
Our results in this context may provide a theoretical framework that
may help interpret and propose new MR experiments to identify the
concentration phases that we see theoretically. This may have impact
on the prediction of the underlying measured geometry. Our results
and analysis can also be of interest in other applications, for example,
in predicting economic growth, and in analyzing the stability of control
systems. 

\section*{Acknowledgment}
We would like to thank Prof. Ofer Pasternak for proposing the idea for the paper. 
\appendix

\section{Transformations of Variable\label{sec:Variable-transformations}}

In this section, we derive the relative magnetization equations, Eq.
(\ref{eq:stochastic Mi}). We use Itô's formula in order to perform the change of variables

\begin{multline*}
df(t,\boldsymbol{m})=\frac{\partial f}{\partial t}dt+\sum_{i}\frac{\partial f}{\partial m_{i}}dm_{i}+\frac{1}{2}\sum_{i,j}\frac{\partial f^{2}}{\partial m_{i}\partial m_{j}}\left[B^{2}\right]_{ij}dt\\
=\left(\frac{\partial f}{\partial t}+\sum_{i}\frac{\partial f}{\partial m_{i}}A_{i}+\frac{1}{2}\sum_{i,j}\frac{\partial f^{2}}{\partial m_{i}\partial m_{j}}\left[B^{2}\right]_{ij}\right)dt+\sum_{ij}\frac{\partial f}{\partial m_{i}}B_{ij}dW_{j},
\end{multline*}
where $A$ and $B$ are the coefficients of the stochastic equations
Eq. (\ref{eq:m0 Ito}) and Eq. (\ref{eq:mi Ito}), respectively, defined
as follows: $A_{i}=J_{i0}h_{0}-J_{0i}m_{i}-\delta\,\underset{j}{\sum}L_{ij}m_{j}+\frac{\sigma_{i}^{2}}{2}m_{i}+f_{i}m_{i}$,
$A_{0}=\underset{j}{\sum}\,J_{0j}m_{j}-\underset{j}{\sum}\,J_{j0}h_{0}+\frac{\sigma_{0}^{2}}{2}h_{0}+f_{0}h_{0}=\frac{J_{\mathrm{in}}}{N}\underset{j}{\sum}\,m_{j}-J_{\mathrm{out}}h_{0}+\frac{\sigma_{0}^{2}}{2}h_{0}+f_{0}h_{0}$,
and $B_{ij}=\delta_{ij}\sigma_{\mathrm{out}}m_{i}$, and $B_{0i}=B_{i0}=\delta_{i0}\sigma_{0}m_{0}$.

\begin{multline*}
\frac{dM_{i}}{dt}=\frac{J_{\mathrm{out}}}{N}-M_{i}\left(\frac{J_{\mathrm{in}}}{N}+J_{\mathrm{in}}\overline{M}-J_{\mathrm{out}}-\frac{\sigma^{2}}{2}+\Delta f\right)-\delta\underset{j}{\sum}L_{ij}M_{j}+\sqrt{\sigma_{\mathrm{out}}^{2}+\sigma_{0}^{2}}\,\xi_{i}M_{i}\\
=\frac{J_{\mathrm{out}}}{N}-M_{i}\left(\frac{J_{\mathrm{in}}}{N}+J_{\mathrm{in}}\overline{M}-J_{\mathrm{out}}-\frac{\sigma^{2}}{2}+\Delta f\right)-\delta\underset{j}{\sum}L_{ij}M_{j}+\sigma\xi_{i}M_{i}\\
=-M_{i}\frac{\sigma^{2}}{2}\left(\beta\overline{M}-\alpha-1-\tilde{\delta}\right)-\delta\overline{M}(t)+\sigma\xi_{i}M_{i}.
\end{multline*}
We introduce the dimensionless parameters $\alpha=2\frac{J_{\mathrm{out}}-\Delta f}{\sigma^{2}}$,
and $\beta=\frac{2J_{\mathrm{in}}}{\sigma^{2}}$. The equations are written under the assumption that the interaction among the nodes and
the hub is described by $J_{i0}=\frac{J_{\mathrm{out}}}{N}$, and $J_{0j}=\frac{J_{\mathrm{in}}}{N}$,
respectively. We also assume, for simplicity, that all the non-hub compartments obey
the statistics $\sigma_{i}=\sigma_{\mathrm{out}}$ and $f_{i}=f$.
The last transition is under the assumption of mean-field interaction $\delta$ among the non-hub nodes. In the Itô form in the limit of $N\rightarrow\infty$, we have
\[
\underset{N\rightarrow\infty}{\mathrm{lim}}\frac{1}{N}\underset{i}{\sum}\sigma_{i}g_{i}(t)M_{i}=0.
\]
Note that one cannot take the limit $N\rightarrow\infty$
in Eq. (\ref{eq:mi Ito}), since the variables $m_{i}$ depend on the
stochastic noise of the hub, $g_{0}$. Taking the sum and letting $N\rightarrow\infty$, we arrive to the following deterministic
equation describing the growth of the average relative mass of the
$N$ non-hub nodes as a function of time:

\begin{equation}
\frac{d\overline{M}(t)}{dt}=\overline{M}(t)\left(J_{\mathrm{out}}+\frac{\sigma^{2}}{2}-\Delta f\right)-J_{\mathrm{in}}\overline{M}(t)^{2}.\label{eq:growth averge m-1}
\end{equation}
The steady-state solutions of this non-linear equation are: 
$\overline{M}_{1\mathrm{eq}}{\rightarrow}-\frac{\Delta f-J_{\mathrm{out}}-\frac{\sigma^{2}}{2}}{J_{\mathrm{in}}}=\frac{1+\alpha}{\beta}$ as ${N\rightarrow\infty}$,
and $\overline{M}_{2\mathrm{eq}}{\rightarrow}0$ as ${N\rightarrow\infty}$.
The equation is also valid when $\delta\neq0$. 

\subsection{Finite-$N$ corrections \label{subsec: finite N mbar}}
In this subsection, we consider finite $N$ corrections to the average equation. With this effect accumulated for the equation reads   

\begin{equation}
\frac{d\overline{M}(t)}{dt}=\varepsilon a +\varepsilon b\overline{M}(t) +c\overline{M}(t)+ b\overline{M}(t)^{2},\label{eq:growth averge m_finite_N}
\end{equation}
where $\varepsilon=\frac{1}{N}$. We denote $a=\frac{\sigma^{2}}{2}J_{\mathrm{out}}$, $b=-\frac{\sigma^{2}}{2}\beta$, and $c=\frac{\sigma^{2}}{2}(\alpha+1)$. The equation has a Riccati form.  Taking the first-order correction in $\varepsilon$, $\overline{M}(t)=\overline{M}_0(t)+\varepsilon \overline{M}_1(t)$, we get 

\begin{equation}
\frac{d\overline{M}_0(t)}{dt}= c\overline{M}_0(t)+ b\overline{M}_0(t)^{2}.\label{eq:growth averge m_0}
\end{equation}
The solution for $\overline{M}_0(t)$ is the logistic function as before. Next,
\begin{equation}
\frac{d\overline{M}_1(t)}{dt}= a + b\overline{M}_0(t)+ (c+2b \overline{M}_0(t)) \overline{M}_1(t).
 \label{eq:growth averge m_1}
\end{equation}
The solution for $\overline{M}_1(t)$ is given by

\begin{multline*}
\overline{M}_1(t) = \overline{M}_1(0)\mathrm{exp}(\int_{0}^{t}(c+2b \overline{M}_0(s))ds)+ \int_{0}^{t}\mathrm{exp}(\int_{s}^{t}(c+2b \overline{M}_0(s))ds)(a + b\overline{M}_0(s))ds
\\=\overline{M}_1(0)\mathrm{exp}(ct+2b\int_{0}^{t} \overline{M}_0(s)ds)+ \int_{0}^{t}\mathrm{exp}(c(t-s)+2b\int_{s}^{t} \overline{M}_0(\tau)d\tau)(a + b\overline{M}_0(s))ds.
\end{multline*}
Substituting the expression of the solution to $\overline{M}_0(t)$, one can show that  $\underset{t\rightarrow\infty}{\mathrm{lim}}\,\overline{M}_{1}(t)$ is finite, meaning that the correction of order $\frac{1}{N}$ to Eq. (\ref{eq:growth averge m-1}) is negligible in the large $N$ limit. 
\section{Moments Lyapunov Exponent\label{sec:Moments-equations}}

In this section, we derive the moment Lyapunov exponent Eq. (\ref{eq:gamma1}).
For this purpose, we present the first moment equations of Eq. (\ref{eq:m0 Ito})
and Eq. (\ref{eq:mi Ito}) for a finite number of non-hub nodes $N$ and $\delta=0$. These equations can be derived using the Fokker-Plank
equation or alternatively the Feynman-Kac formula \cite{seroussi2018spectral,schuss2009theory,molchanov1991ideas,carmona1994parabolic}.
They read

\begin{equation}
\frac{d\langle h_{0}\rangle}{dt}=\frac{J_{\mathrm{in}}}{N}\langle\overline{m}\rangle-J_{\mathrm{out}}\langle h_{0}\rangle+f_{0}\langle h_{0}\rangle+\frac{\sigma_{0}^{2}}{2}\langle h_{0}\rangle,\label{eq:first moment m0}
\end{equation}

\begin{equation}
\frac{d\langle m_{i}\rangle}{dt}=\frac{J_{\mathrm{out}}}{N}\langle h_{0}\rangle-\frac{J_{\mathrm{in}}}{N}\langle m_{i}\rangle+f\langle m_{i}\rangle+\frac{\sigma_{\mathrm{out}}^{2}}{2}\langle m_{i}\rangle.\label{eq:first moment mi}
\end{equation}
Here it is assumed that all the non-hub nodes are independent and with
the same dynamics and denoting the total mass, $\langle\overline{m}\rangle=N\langle m_{i}\rangle$.
Summing over $N$ in Eq. (\ref{eq:first moment mi}), we get

\begin{equation}
\frac{d\langle\overline{m}\rangle}{dt}=J_{\mathrm{out}}\langle h_{0}\rangle-\frac{J_{\mathrm{in}}}{N}\langle\overline{m}\rangle+f\langle\overline{m}\rangle+\frac{\sigma_{\mathrm{out}}^{2}}{2}\langle\overline{m}\rangle.\label{eq:first moment mbar}
\end{equation}
Eqs. (\ref{eq:first moment mbar}) and (\ref{eq:first moment m0}) can be written in vector form as
\[
\frac{d\boldsymbol{a}}{dt}=A\boldsymbol{a}(t)=\left(\begin{array}{cc}
f_{0}-J_{\mathrm{out}}+\frac{\sigma_{0}^{2}}{2} & \frac{J_{\mathrm{in}}}{N}\\
J_{\mathrm{out}} & f-\frac{J_{\mathrm{in}}}{N}+\frac{\sigma_{\mathrm{out}}^{2}}{2}
\end{array}\right)\boldsymbol{a}(t).
\]
This is a simple system of linear equations, and the dynamics it describes is determined by the eigenvalues of the matrix $A$. The resulted
eigenvalues are than,

\begin{equation}
\lambda_{1}=f+\frac{\sigma_{\mathrm{out}}^{2}}{2}-\frac{2J_{\mathrm{in}}}{N}\left(1+\frac{J_{\mathrm{out}}}{(\Delta f+\frac{\Delta\sigma^{2}}{2}-\frac{J_{\mathrm{in}}}{N}-J_{\mathrm{out}})}\right),
\end{equation}
and, 
\begin{equation}
\lambda_{2}=f_{0}-J_{\mathrm{out}}+\frac{\sigma_{0}^{2}}{2}+\frac{2J_{\mathrm{out}}\frac{J_{\mathrm{in}}}{N}}{(\Delta f+\frac{\Delta\sigma^{2}}{2}-\frac{J_{\mathrm{in}}}{N}-J_{\mathrm{out}})}.
\end{equation}
The solution is then given by $\left(\begin{array}{c}
\langle m_{0}\rangle(t)\\
\langle\overline{m}\rangle(t)
\end{array}\right)=Av_{1}e^{\lambda_{1}t}+Bv_{2}e^{\lambda_{2}t}=A\left(\begin{array}{c}
\lambda_{1}-d\\
c
\end{array}\right)e^{\lambda_{1}t}+B\left(\begin{array}{c}
\lambda_{2}-d\\
c
\end{array}\right)e^{\lambda_{2}t}=A\left(\begin{array}{c}
\Delta f-J_{\mathrm{out}}+\frac{\Delta\sigma^{2}}{2}\\
J_{\mathrm{out}}
\end{array}\right)e^{\lambda_{1}t}+B\left(\begin{array}{c}
0\\
J_{\mathrm{out}}
\end{array}\right)e^{\lambda_{2}t}$. Plugging this expression into the equation of the first moment Lyapunov
exponent we get

\begin{multline*}
\gamma_{1}=\underset{t\rightarrow\infty}{\mathrm{lim}}\,\underset{N\rightarrow\infty}{\mathrm{lim}}\,\frac{1}{t}\mathrm{ln}\left(\langle h_{0}\rangle+\langle\overline{m}\rangle\right)
\\=\mathrm{\underset{t\rightarrow\infty}{lim}}\,\frac{1}{t}\mathrm{ln}\,e^{\lambda_{2}t}\left(A\left(\Delta f-J_{0}+\frac{\Delta\sigma^{2}}{2}+J_{\mathrm{out}}\right)e^{\left(\lambda_{1}-\lambda_{2}\right)t}+BJ_{\mathrm{out}}\right)\\
=\begin{cases}
\begin{array}{c}
f+\frac{\sigma_{\mathrm{out}}^{2}}{2}\\
f_{0}-J_{\mathrm{out}}+\frac{\sigma_{0}^{2}}{2}
\end{array} & \begin{array}{c}
\frac{\Delta\sigma^{2}}{2}+\Delta f-J_{\mathrm{out}}<0\\
\frac{\Delta\sigma^{2}}{2}+\Delta f-J_{\mathrm{out}}\geq0
\end{array}\end{cases}\\=\begin{cases}
\begin{array}{c}
f+\frac{\sigma_{\mathrm{out}}^{2}}{2}\\
f_{0}-J_{\mathrm{out}}+\frac{\sigma_{0}^{2}}{2}
\end{array} & \begin{array}{c}
\alpha>\frac{\Delta\sigma^{2}}{\sigma^{2}}\\
\alpha\leq\frac{\Delta\sigma^{2}}{\sigma^{2}}
\end{array}\end{cases}.
\end{multline*}

\section{Sample Lyapunov Exponent\label{sec:Sample-Lyapunov-Exponent}}

Here, we calculate the sample Lyapunov exponent of the
mass of the system, defined as 

\[
\widetilde{\gamma}=\underset{t,N\rightarrow\infty}{\mathrm{lim}}\,\frac{1}{t}\mathrm{ln}\left(h_{0}+\sum_{i}m_{i}\right).
\]
To prove the formula (\ref{eq:sample Lyapunov}) in the main text,
we consider at the system of equations

\[
\frac{dm_{i}}{dt}=\frac{J_{\mathrm{out}}}{N}h_{0}-\frac{J_{\mathrm{in}}}{N}m_{i}+fm_{i}+\frac{\sigma_{\mathrm{out}}^{2}}{2}m_{i}+\sigma_{\mathrm{out}}g_{i}(t)m_{i},
\]

\[
\frac{dh_{0}}{dt}=\frac{J_{\mathrm{in}}}{N}\overline{m}+\left(f_{0}+\frac{\sigma_{0}^{2}}{2}-J_{\mathrm{out}}\right)h_{0}+\sigma_{0}g_{0}(t)h_{0}.
\]
Using a transformation of variable given by the Itô formula

\begin{multline*}
\frac{d\mathrm{ln}\left(\overline{m}+h_{0}\right)}{dt}=\frac{\partial f}{\partial t}+\sum_{i}\frac{\partial f}{\partial m_{i}}A_{i}+\frac{\partial f}{\partial h_{0}}A_{0}+\frac{1}{2}\frac{\partial f^{2}}{\partial^{2}h_{0}}\left[B^{2}\right]_{00}\\
+\frac{1}{2}\sum_{i}\frac{\partial f^{2}}{\partial^{2}m_{i}}\left[B^{2}\right]_{ii}+\frac{\partial f}{\partial h_{0}}B_{00}\frac{dW_{0}}{dt}+\sum_{i}\frac{\partial f}{\partial m_{i}}B_{ii}\frac{dW_{i}}{dt}\\
=\frac{1}{N\overline{M}+1}\left(J_{\mathrm{out}}-J_{\mathrm{in}}\overline{M}+f\overline{M}+\frac{\sigma_{\mathrm{out}}^{2}}{2}\overline{M}+\sigma_{\mathrm{out}}\sum_{i}g_{i}(t)M_{i}\right)\\
+\frac{1}{N\overline{M}+1}\left(J_{\mathrm{in}}\overline{M}+f_{0}+\frac{\sigma_{0}^{2}}{2}-J_{\mathrm{out}}\right)-\frac{1}{2}\frac{\sigma_{0}^{2}+\sigma_{\mathrm{out}}^{2}\sum_{i}M_{i}^{2}}{\left(N\overline{M}+1\right)^{2}}\\
+\frac{1}{N\overline{M}+1}\sigma_{0}g_{0}(t)+\sigma_{\mathrm{out}}\sum_{i}\frac{g_{i}(t)M_{i}}{N\overline{M}+1}\\
=\frac{1}{N\overline{M}+1}\sigma_{0}g_{0}(t)+\sigma_{\mathrm{out}}\sum_{i}\frac{g_{i}(t)M_{i}}{N\overline{M}+1}\\
+\frac{1}{N\overline{M}+1}\left(\left(f+\frac{\sigma_{\mathrm{out}}^{2}}{2}\right)N\overline{M}+\left(f_{0}+\frac{\sigma_{0}^{2}}{2}\right)\right)-\frac{1}{2}\frac{\sigma_{0}^{2}+\sigma_{\mathrm{out}}^{2}\sum_{i}M_{i}^{2}}{\left(N\overline{M}+1\right)^{2}}.
\end{multline*}
Integrating with respect to time and dividing by $t$, we have

\[
\frac{1}{t}\mathrm{ln}\left(\overline{m}(t)+h_{0}(t)\right)=\frac{1}{t}\int_{0}^{t}\frac{d\mathrm{ln}\left(\overline{m}(\tau)+h_{0}(\tau)\right)}{d\tau}d\tau.
\]
Letting the limit of $t,N\rightarrow\infty$, we get

\begin{multline}
\widetilde{\gamma}=\underset{t,N\rightarrow\infty}{\mathrm{lim}}\,\frac{1}{t}\mathrm{ln}\left(\overline{m}+h_{0}\right)=\underset{t, N\rightarrow\infty}{\mathrm{lim}}\frac{1}{t}\int_{0}^{t}\frac{d\mathrm{ln}\left(\overline{m}(\tau)+h_{0}(\tau)\right)}{d\tau}d\tau\\
=\underset{t,N\rightarrow\infty}{\mathrm{lim}}\,\frac{1}{t}\int_{0}^{t}\left[\frac{\sigma_{0}g_{0}(\tau)}{N\overline{M}+1}-\frac{1}{2}\frac{\sigma_{0}^{2}}{\left(N\overline{M}+1\right)^{2}}+\sigma_{\mathrm{out}}\sum_{i}\frac{g_{i}(\tau)M_{i}}{N\overline{M}+1}-\frac{\sigma_{\mathrm{out}}^{2}}{2}\frac{\sum_{i}M_{i}^{2}}{\left(N\overline{M}+1\right)^{2}}\right]d\tau\\
+\underset{t,N\rightarrow\infty}{\mathrm{lim}}\,\frac{1}{t}\int_{0}^{t}\frac{1}{N\overline{M}+1}\left(\left(f+\frac{\sigma_{\mathrm{out}}^{2}}{2}\right)N\overline{M}+\left(f_{0}+\frac{\sigma_{0}^{2}}{2}\right)\right)d\tau.\label{eq:smaple detailed integrals}
\end{multline}
Here we used the deterministic law of $\overline{M}$ given that higher corrections in $N$ are negligible, see \ref{subsec: finite N mbar} for more details. Using the stationarity property of the Gaussian processes $g_{0}(t)$
and $g_{i}(t)$, and the ergodic theorem: 
\begin{multline}
\widetilde{\gamma}=\underset{t,N\rightarrow\infty}{\mathrm{lim}}\,\frac{1}{t}\int_{0}^{t}\left[\frac{1}{\left(N\overline{M}+1\right)}\left(\left(f+\frac{\sigma_{\mathrm{out}}^{2}}{2}\right)N\overline{M}+\left(f_{0}+\frac{\sigma_{0}^{2}}{2}\right)\right)\right]d\tau\\
=\underset{t,N\rightarrow\infty}{\mathrm{lim}}\left(f+\frac{\sigma_{\mathrm{out}}^{2}}{2}\right)\frac{1}{t}\int_{0}^{t}\frac{\overline{M}}{\overline{M}+\frac{1}{N}}d\tau+\left(f_{0}+\frac{\sigma_{0}^{2}}{2}\right)\frac{1}{t}\int_{0}^{t}\frac{d\tau}{N\overline{M}+1}.\label{Eq:gmSample2terms}
\end{multline}

This formula is obtained under the assumption that the fluctuations of the noise are
finite (Novikov condition), i.e., 

\begin{equation}
\underset{t,N\rightarrow\infty}{\mathrm{lim}}\,\frac{1}{t}\int_{0}^{t}\frac{\sum_{i}M_{i}^{2}}{\left(N\overline{M}+1\right)^{2}}d\tau<\infty\label{eq:fluctuation 1}
\end{equation}
and, 
\begin{equation}
\underset{t,N\rightarrow\infty}{\mathrm{lim}}\,\frac{1}{t}\int_{0}^{t}\frac{1}{\left(N\overline{M}+1\right)^{2}}d\tau<\infty.\label{eq:fluctuation 2}
\end{equation}
In order to calculate the above integrals one needs to specify how the number of nodes in the graphs growth with time. We show that if the number of nodes grows exponentially in $t$, i.e., $N\sim e^{\varepsilon t}$, then there exists a localization phase. In addition, the fluctuation, conditions \ref{eq:fluctuation 1} and \ref{eq:fluctuation 2}, are satisfied. These conditions are verified in \ref{subsec:Noise-fluctuations}
below. Since the fluctuations are finite, we are left with calculating integrals
of the logistic function $\overline{M}(t)$. We use the following
properties of the logistic function:

\begin{equation}
\overline{M}(t)=\frac{1}{A+Be^{-\xi t}}=\frac{e^{\xi t}}{Ae^{\xi t}+B}\label{eq:logistic identety1}
\end{equation}
and
\begin{equation}
\int_{0}^{t}\overline{M}d\tau=\frac{1}{\xi A}\mathrm{log}\left(Ae^{\xi t}+B\right).\label{eq:logistic identety2}
\end{equation}
In our case, $A=\frac{\beta}{\alpha+1}$, $B=\frac{1}{\overline{M}_{0}}-\frac{\beta}{\alpha+1}$,
and $\xi=\frac{\sigma^{2}\left(\alpha+1\right)}{2}$. 
Using Eqs. (\ref{eq:logistic identety1}),
and (\ref{eq:logistic identety2}) it is easy to show that
\begin{multline}
\underset{t,N\rightarrow\infty}{\mathrm{lim}}\,\frac{1}{t}\int_{0}^{t}\frac{\overline{M}}{\overline{M}+\frac{1}{N}}d\tau=\underset{t\rightarrow\infty}{\mathrm{lim}}\,\frac{1}{t}\int_{T}^{t}\frac{\overline{M}}{\overline{M}+e^{-\varepsilon(\tau-T)}}d\tau
\\=\underset{t\rightarrow\infty}{\mathrm{lim}}\frac{1}{t}\int_{T}^{t}\frac{1}{1+Ae^{\varepsilon T}e^{-\varepsilon\tau}+Be^{\varepsilon T}e^{-\left(\xi+\varepsilon\right)\tau}}d\tau=1,\label{eq:term1}
\end{multline}
where $T$ is some finite time. In the same manner, one can calculate the second term in Eq. (\ref{Eq:gmSample2terms}):

\begin{multline}
\underset{t,N\rightarrow\infty}{\mathrm{lim}}\,\frac{1}{t}\int_{0}^{t}\frac{d\tau}{N\overline{M}+1}=\underset{t\rightarrow\infty}{\mathrm{lim}}\frac{1}{t}\int_{0}^{t}\frac{A+Be^{-\xi\tau}d\tau}{N+A+Be^{-\xi\tau}}\\=\underset{t\rightarrow\infty}{\mathrm{lim}}\frac{1}{t}\int_{0}^{t}\frac{Ad\tau}{e^{\varepsilon(\tau-T)}+A+Be^{-\xi\tau}}+\underset{t\rightarrow\infty}{\mathrm{lim}}\,\frac{1}{t}\int_{0}^{t}\frac{Be^{-\xi\tau}}{e^{\varepsilon(\tau-T)}+A+Be^{-\xi\tau}}d\tau\\=\underset{t\rightarrow\infty}{\mathrm{lim}}\,\frac{1}{t}\int_{0}^{t}\frac{Be^{-\left(\xi+\varepsilon\right)\tau}}{e^{-\varepsilon T}+Be^{-\left(\xi+\varepsilon\right)\tau}}d\tau=\underset{t\rightarrow\infty}{\mathrm{lim}}\,-\frac{1}{t\left(\xi+\varepsilon\right)}\mathrm{log}\left(e^{-\varepsilon T}+Be^{-\left(\xi+\varepsilon\right)t}\right)=0,\label{eq:term2}
\end{multline}
provided that $\mathrm{max}\{-\xi,0\}\leq\varepsilon$.  Note that, in order to have finite fluctuations, $\sigma^{2}\leq\epsilon$,
see \ref{subsec:Noise-fluctuations}. Substituting the results
in Eq. (\ref{eq:term1}) and Eq. (\ref{eq:term2}) we obtain 
\begin{equation}
\widetilde{\gamma}=f+\frac{\sigma_{\mathrm{out}}^{2}}{2}.
\end{equation}
Therefore, for any exponential growth rate $\sigma^{2}\mathrm{max}\{-\frac{\alpha+1}{2},1\}\leq\varepsilon$, the fluctuations and sample Lyapunov are finite.

\subsection{Noise fluctuations\label{subsec:Noise-fluctuations} }

In this subsection, we prove that the fluctuation are finite when $N\sim e^{\varepsilon t}$,
i.e., the conditions in (\ref{eq:fluctuation 1}) and (\ref{eq:fluctuation 2})
are satisfied. That condition (\ref{eq:fluctuation 2}) is satisfied, is readily seen since the function under the integral is bounded between zero and one, and so the integral itself is also finite: 

\begin{equation*}
\underset{t,N\rightarrow\infty}{\mathrm{lim}}\,\frac{1}{t}\int_{0}^{t}\frac{d\tau}{\left(N\overline{M}+1\right)^{2}}
<\infty\label{eq:Fluctuations hub}
\end{equation*}
In order to calculate the fluctuation of the non-hub nodes, i.e., establish (\ref{eq:fluctuation 1}), we approximate the limit, $\underset{N\rightarrow\infty}{\mathrm{lim}}\frac{1}{N}\sum_{i}M_{i}^{2}\rightarrow\langle M_{i}^{2}\rangle$;
since $M_{i}$ are i.i.d. random variables. The
second moment can be calculated using the Fokker-Plank equation Eq.
(\ref{eq:FP M0}):
\[
\frac{\partial P(M,t)}{\partial t}=-\frac{\sigma^{2}}{2}\frac{\partial\left(\left(\left(\alpha+1-\beta\overline{M}\right)M\right)P(M,t)\right)}{\partial M}+\frac{\sigma^{2}}{2}\frac{\partial^{2}\left(M^{2}P(M,t)\right)}{\partial^{2}M}.
\]
Averaging over the second moment yields:
\[
\frac{d\langle M^{2}\rangle}{dt}=\frac{\sigma^{2}}{2}\left(2\left(\alpha+2-\beta\overline{M}\right)\langle M^{2}\rangle\right).
\]
The solution for $\tilde{\delta}=0$ is 
\begin{multline*}
\langle M^{2}(t)\rangle=\langle M^{2}\rangle(0)\mathrm{exp}\left(\sigma^{2}t\left(\alpha+2\right)-\sigma^{2}\beta\int_{0}^{t}\overline{M}d\tau\right)\\=\frac{\langle M^{2}\rangle(0)e^{\sigma^{2}t}}{\left(\frac{\beta}{(\alpha+1)}+\frac{\beta\left(\frac{\alpha+1}{\beta}-\overline{M}_{0}\right)}{\overline{M}_{0}(\alpha+1)}e^{-\frac{\sigma^{2}\left(\alpha+1\right)t}{2}}\right)^{2}}
=\langle M^{2}\rangle(0)e^{\sigma^{2}t}\overline{M}(t)^{2}.
\end{multline*}
The fluctuations of the non-hub nodes, i.e., the last term in Eq. (\ref{eq:smaple detailed integrals})
are then,  given by
\begin{multline*}
\underset{t,N\rightarrow\infty}{\mathrm{lim}}\,\frac{1}{t}\int_{0}^{t}\frac{\langle M_{i}^{2}\rangle}{N\left(\overline{M}+\frac{1}{N}\right)^{2}}d\tau=\underset{t,N\rightarrow\infty}{\mathrm{lim}}\,\frac{1}{t}\int_{0}^{t}\frac{\langle M_{i}^{2}\rangle}{N\left(\overline{M}+\frac{1}{N}\right)^{2}}d\tau\\
=\langle M^{2}\rangle(0)\underset{t,N\rightarrow\infty}{\mathrm{lim}}\,\frac{1}{t}\int_{0}^{t}\frac{e^{\sigma^{2}t}\overline{M}^{2}}{N\left(\overline{M}+\frac{1}{N}\right)^{2}}d\tau.
\end{multline*}
Substituting here the logistic function $\overline{M}(t)$ (Eq. (\ref{eq:logistic identety1})), we get

\begin{multline*}
\underset{t,N\rightarrow\infty}{\mathrm{lim}}\frac{1}{t}\int_{0}^{t}\frac{e^{\sigma^{2}\tau}\overline{M}^{2}}{N\left(\overline{M}+\frac{1}{N}\right)^{2}}d\tau=\underset{t,N\rightarrow\infty}{\mathrm{lim}}\frac{1}{t}\int_{0}^{t}\frac{e^{\sigma^{2}\tau}}{N+2\left(A+Be^{-\xi\tau}\right)+\frac{1}{N}\left(A+Be^{-\xi\tau}\right)^{2}}d\tau\\=\underset{t\rightarrow\infty}{\mathrm{lim}}\frac{1}{t}\int_{0}^{t}\frac{e^{\sigma^{2}\tau}}{e^{\varepsilon(\tau-T)}+2\left(A+Be^{-\xi\tau}\right)+e^{-\varepsilon(\tau-T)}\left(A+Be^{-\xi\tau}\right)^{2}}d\tau\\=\underset{t\rightarrow\infty}{\mathrm{lim}}\frac{1}{t\left(\sigma^{2}-\varepsilon\right)}e^{\left(\sigma^{2}-\varepsilon\right)\left(t-T\right)+\varepsilon T}d\tau=0.\label{eq:fluctuations small nodes}
\end{multline*}
Therefore, the fluctuation are finite for any $\varepsilon\geq\sigma^2$. 

\section{Multiple Hubs Derivation\label{sec:Multiple-Hubs-Derivation}}

In this section, we derive the normalized equation in the multiple
hubs models of Sec. \ref{Sec: multipileHaubs}. Taking normalized variables
$M_{i}=\frac{m_{i}}{\overline{h}}$, $x_{i}=\frac{h_{i}}{\overline{h}}$,
so that, $\overline{M}=\frac{1}{N}\sum_{i}\frac{m_{i}}{\overline{h}}$,
and $\frac{1}{H}\sum_{j}x_{j}=1$, Eq. (\ref{eq:hi-multi-hub}), and
Eq. (\ref{eq:mi-multi-hub}) are transformed into the following set
of equations:

\begin{equation}
\frac{dM_{i}}{dt}=\frac{J_{\mathrm{out}}}{N}-\left(\frac{J_{\mathrm{in}}}{N}-\frac{J_{\mathrm{out}}}{H}-\Delta f+\frac{J_{\mathrm{in}}}{H}\overline{M}\right)M_{i}+\frac{\nu_{i}^{2}\left(x\right)}{2}M_{i}+\nu_{i}\left(x\right)M_{i}\xi_{i}\label{eq:Mi-multi-hub}
\end{equation}

\begin{equation}
\frac{dx_{i}}{dt}=J_{\mathrm{in}}\overline{M}\left(1-x_{i}\right)+\frac{\sigma_{i}^{2}\left(x\right)}{2}x_{i}+\sigma_{i}\left(x\right)x_{i}\phi_{i}.\label{eq:xi-Multi-hub}
\end{equation}
Set 
\begin{equation}
 \nu_{i}f_{i}(t)-\frac{1}{H}\sum_{j}^{H}\sigma_{j}g_{j}(t)x_{j}=\sqrt{\nu_{i}^{2}+\frac{1}{H}\sum_{j}^{H}\sigma_{j}^{2}x_{j}^{2}}\xi_{i}=\nu_{i}\left(x\right)\xi_{i}
\end{equation}
and
\begin{equation} \sigma_{i}g_{i}(t)-\frac{1}{H}\sum_{j}^{H}\sigma_{j}g_{j}(t)x_{j}=\sqrt{\sigma_{i}^{2}+\frac{1}{H}\sum_{j}^{H}\sigma_{j}^{2}x_{j}^{2}}\phi_{i}=\sigma_{i}\left(x\right)\phi_{i},
\end{equation}
so that $\frac{1}{H}\sum_{i}^{H}\sigma_{i}\left(x\right)\phi_{i}x_{i}=0$,
and $\nu_{i}^{2}\left(x\right)-\nu_{i}^{2}=\sigma_{i}^{2}\left(x\right)-\sigma_{i}^{2}$,
and $\Delta f=f-q$. In the limit of $H,N\rightarrow\infty$, the
variances are constants, $\nu^{2}\left(x\right)=\nu^{2}+\sigma^{2}$,
and $\sigma^{2}\left(x\right)=2\sigma^{2}$. In this limit,one can average Eq. (\ref{eq:Mi-multi-hub}), since the variables are decoupled.
\bibliography{bibGiantComponent}
\end{document}